\documentstyle[epsfig,prb,aps,multicol]{revtex}
\begin{document}
\draft
\title{Magnetic subband structure of an electron on a triangular lattice}
\author{Gi-Yeong Oh$^{*}$}
\address{Department of Basic Science, Hankyong National University,}
\address{Kyonggi-do 456-749, Korea}
\date{submitted to Phys. Rev. B on June 29, 1999}
\maketitle

\begin{abstract}
The effects of anisotropic nearest-neighbor (NN) and isotropic
next-nearest-neighbor (NNN) hopping integrals on the magnetic subband
structure of an electron on a triangular lattice are studied within the
tight-binding approximation. A generalized Harper equation that
includes both the NN and the NNN hopping terms is first derived. Then,
the effects of anisotropic NN hopping integrals are examined and the
results are compared in detail with those of an existing literature
[Phys. Rev. B {\bf 56}, 3787 (1997)]. It is found that the hopping
anisotropy generically leads to the occurrence of band broadening and
gap closing. The effects of isotropic NNN hopping integrals are next
examined, and it is found that introducing the NNN hopping integral
changes considerably the magnetic subband structure; band broadening,
gap closing, gap reopening, and band crossing occur depending on the
strength of the isotropic NNN hopping integral. Symmetries of the
magnetic subband structures with and without both the hopping
anisotropy and the isotropic NNN hopping integrals are also discussed.

\end{abstract}

\begin{multicols}{2}

\section{Introduction}
The problem of a two-dimensional (2D) Bloch electron under a
perpendicular magnetic field has attracted much interest for several
decades, since not only it exhibits extremely rich energy band
structures\cite{1}$^{-}$\cite{7} but also it is closely related with
various phenomena such as the quantum Hall effect,\cite{a} the
flux-state model for high-$T_{c}$ superconductivity,\cite{b} and the
mean-field transition temperature of superconducting networks or
Josephson junction arrays.\cite{c} Recently, by means of the remarkable
advent of nanofabrication techniques, studies of finding the indication
of the peculiar band structure and its effect on the transport and
optical properties have been also extensively
performed.\cite{8}$^{-}$\cite{11}

A traditional method to solve this problem is to start with energy
dispersion $\varepsilon(\vec{k})$ without a magnetic field and then
make the Peierls substitution $\vec{k}\rightarrow(\vec{p}+
e\vec{A})/\hbar$ to construct an effective single-band Hamiltonian
\begin{equation}
 H=\varepsilon[(\vec{p}+e\vec{A})/\hbar],             \label{eq:0}
\end{equation}
where $\vec{p}$ is the momentum operator and $\vec{A}$ is a vector
potential. An equivalent way to this method is to start directly with
the tight-binding Hamiltonian given by
\begin{equation}
 H=\sum_{ij}t_{ij}e^{i\theta_{ij}}|i\rangle\langle j|, \label{eq:1}
\end{equation}
where $t_{ij}$ is the hopping integral between the sites $i$ and $j$,
$|i\rangle$ is a state of an atomiclike orbital centered at the site
$i$, and $\theta_{ij}$ is the magnetic phase factor defined by
\begin{equation}
 \theta_{ij}=\frac{2\pi}{\phi_{0}}\int_{i}^{j}\vec{A}\cdot d\vec{l},
                                                       \label{eq:2}
\end{equation}
$\phi_{0}=hc/e$ being the magnetic flux quantum.

So far, a number of works have focused on the energy spectrum of a
lattice with a square symmetry, and the magnetic subband structure of
the lattice with isotropic hopping integrals between nearest-neighbor
(NN) sites is well known as the Hofstadter's butterfly.\cite{2,4} The
effect of the anisotropy in the NN hopping integrals on the magnetic
subband structure was also studied, and it was found that introducing
the hopping anisotropy leads to the occurrence of band broadening and
gap closing.\cite{12}$^{-}$\cite{15} Furthermore, the effect of
next-nearest-neighbor (NNN) hopping integrals on the magnetic subband
structure was also studied,\cite{16} and introducing the NNN hopping
integrals was found to play a key role in removing the degeneracy which
appears at the band center of the energy spectrum under isotropic NN
hopping integrals.

Along with the case of the square lattice, magnetic subband structures
of a triangular lattice have been intensively
studied,\cite{13,17}$^{-}$\cite{21} and it was found that the lattice
with isotropic NN hopping integrals exhibits a recursive magnetic
subband structure similarly to the case of the square lattice with
isotropic NN hopping integrals.\cite{18} Recently, the effect of the
anisotropy in the NN hopping integrals on the magnetic subband
structure was studied by Gumbs and Fekete.\cite{21} In studying the
effect, they applied the Peierls substitution for the energy dispersion
given by
\begin{eqnarray}
 \varepsilon(\vec{k})=&&2\left\{t_{a}\cos(k_{x}a)+t_{b}\cos\left[(k_{x}
 +\sqrt{3}k_{y}) \frac{a}{2}\right]\right.\nonumber\\&&\left.+t_{c}\cos
 \left[(k_{x}-\sqrt{3}k_{y}) \frac{a}{2}\right]\right\} \label{eq:10}
\end{eqnarray}
to construct a Hamiltonian matrix. And, by diagonalizing the resultant
matrix, they obtained magnetic subband structures for a few sets of
($t_{a},t_{b},t_{c}$) and presented relevant arguments. However,
unfortunately, most of the results and arguments presented by them are
erroneous ones that arise from a mistake in deriving the Hamiltonian,
as we shall show below. We re-examine the same problem as that
investigated by Gumbs and Fekete to put the correct formalism and
numerical results, which is the first end of this paper. The second end
of this paper is to examine the effect of isotropic NNN hopping
integrals on the magnetic subband structure. To our knowledge, few work
has been devoted to this problem. However, noting that the triangular
lattice with NNN hopping integrals is topologically equivalent to the
square lattice with NNN hopping integrals, it may be interesting to
examine this problem and compare the results with the case of the
square lattice. We will show that the magnetic subband structure
changes considerably depending on the strength of the isotropic NNN
hopping integrals.

This paper is organized as follows: In Sec.~II, we derive a generalized
Harper equation that includes NNN as well as NN hopping integrals. In
Sec.~III, we first point out the mistake and the misleading arguments
presented in Ref.~24, and then present correct numerical results on the
effects of both anisotropic NN and isotropic NNN hopping integrals on
the magnetic subband structure. We also discuss the symmetries of the
magnetic subband structure obtained from the generalized Harper
equation. Finally, Sec.~IV is devoted to a brief summary.

\section{The tight-biding equation}

We consider an electron on a 2D triangular lattice with a hexagonal
symmetry in the presence of a uniform perpendicular magnetic field
$\vec{B}=B\hat{z}$. Under the Landau gauge, the vector potential is
given by $\vec{A}=(0,Bx,0)$ and $\theta_{ij}$ is given by
\begin{equation}
 \theta_{ij}= \left\{ \begin{array}{ll}
  0, &~~ j=(m\pm2,n)\\
  \pm\pi\phi(m\pm 1/2), &~~ j=(m\pm1,n\pm1)\\
  \pm\pi\phi(m\mp 1/2), &~~ j=(m\mp1,n\pm1)\\
  \pm\pi\phi(m\pm 3/2), &~~ j=(m\pm3,n\pm1)\\
  \pm\pi\phi(m\mp 3/2), &~~ j=(m\mp3,n\pm1)\\
  \pm2\pi\phi m, &~~ j=(m,n\pm2)
  \end{array}\right.,                                  \label{eq:3}
\end{equation}
where $i=(m,n)$. Here, $(m,n)$ labels the lattice points, i.e.,
$(x,y)=(mb,nc)$, where $b=a/2$, $c=\sqrt{3}a/2$, and $a$ is the lattice
constant. And, $\phi=2\sqrt{3}Bb^{2}/\phi_{0}$ is the magnetic flux
through the unit cell. By means of Eqs.~(\ref{eq:1}) and (\ref{eq:3}),
the tight-binding equation, $H\Psi_{m,n}=E\Psi_{m,n}$, can be written
as
\begin{eqnarray}
 E&&\Psi_{m,n}=t_{a}(\Psi_{m-2,n}+\Psi_{m+2,n}) \nonumber \\
   &&+t_{b}[e^{i\pi\phi(m-1/2)}\Psi_{m-1,n-1}
     +e^{-i\pi\phi(m+1/2)}\Psi_{m+1,n+1}] \nonumber \\
   &&+t_{c}[e^{i\pi\phi(m+1/2)}\Psi_{m+1,n-1}
     +e^{-i\pi\phi(m-1/2)}\Psi_{m-1,n+1}]  \nonumber \\
   &&+t_{a}'[e^{i\pi\phi(m-3/2)}\Psi_{m-3,n-1}
     +e^{-i\pi\phi(m+3/2)}\Psi_{m+3,n+1}] \nonumber \\
   &&+t_{c}'[e^{i\pi\phi(m+3/2)}\Psi_{m+3,n-1}
     +e^{-i\pi\phi(m-3/2)}\Psi_{m-3,n+1}]  \nonumber \\
   &&+t_{b}'(e^{i2\pi\phi m}\Psi_{m,n-2}
     +e^{-i2\pi\phi m}\Psi_{m,n+2}) ,          \label{eq:4}
\end{eqnarray}
where $t_{a}$ is the NN hopping integral along the $x$ direction,
$t_{b (c)}$ is the NN hopping integral along the direction
which makes $\pi/3$ ($2\pi/3$) with the $x$ direction,
and $t_{a (b,c)}'$ is the NNN hopping integral along the
direction which makes $\pi/6$ ($\pi/2$, $5\pi/6$) with the $x$
direction, respectively.

Since Eq.~(\ref{eq:4}) is invariant under the transformation
$y\rightarrow y+\sqrt{3}a$, $\Psi_{m,n}$ can be written as
\begin{equation}
 \Psi_{m,n}=e^{i\sqrt{3}k_{y}na}\psi_{m}.         \label{eq:bc}
\end{equation}
Thus, using Eqs.~(\ref{eq:4}) and (\ref{eq:bc}), we obtain a kind of
generalized Harper equation as follows:
\begin{eqnarray}
 E\psi_{m}=&&C_{m-2}^{*}\psi_{m-3}+t_{a}\psi_{m-2}
           +B_{m-1}^{*}\psi_{m-1}+A_{m}\psi_{m} \nonumber \\
           &&+B_{m}\psi_{m+1}+t_{a}\psi_{m+2}
           +C_{m+1}\psi_{m+3},                      \label{eq:5}
\end{eqnarray}
where
\begin{eqnarray}
 A_{m}&=&2t_{b}'\cos(2\theta_{m}-\pi\phi),  \nonumber \\
 B_{m}&=&t_{c}e^{i\theta_{m}}+t_{b}e^{-i\theta_{m}}, \nonumber \\
 C_{m}&=&t_{c}' e^{i\theta_{m}}+t_{a}' e^{-i\theta_{m}}, \label{eq:6}
\end{eqnarray}
with $\theta_{m}=\pi\phi(m-1/2)-k_{y}c$. Denoting $\phi=p/q$ with
relative primes $p$ and $q$, it can be easily checked that
\begin{equation}
 A_{m+q}=A_{m},~ B_{m+M}=B_{m},~ C_{m+M}=C_{m},     \label{eq:7}
\end{equation}
where $M=q$ ($2q$) for an even (odd) $p$. Thus, $m$ in Eq.~(\ref{eq:5})
satisfies the condition $1\le m\le q$ ($1\le m\le 2q$) for an even
(odd) $p$, and the Bloch condition along the $x$ direction can be
written as
\begin{equation}
  \psi_{m+M}=e^{ik_{x}Mb}\psi_{m} .                 \label{eq:8}
\end{equation}
Using Eqs.~(\ref{eq:5}) and (\ref{eq:8}), we obtain the eigenvalue
equation, ${\bf A}\Psi=E\Psi$, where
$\Psi^{\dagger}=(\psi_{1},\psi_{2},\cdots,\psi_{M})$ and
\end{multicols}

\begin{equation}
 {\bf A}=\left( \begin{array}{cccccccc}
  A_{1} & B_{1} & t_{a} & C_{2} & \cdots & C_{M-1}^{*}e^{-i\delta} &
    t_{a}e^{-i\delta} & B_{M}^{*}e^{-i\delta} \\
  B_{1}^{*} & A_{2} & B_{2} & t_{a} & \cdots & 0 &
    C_{M}^{*}e^{-i\delta} & t_{a}e^{-i\delta} \\
  t_{a} & B_{2}^{*} & A_{3} & B_{3} & \cdots & 0 & 0 &
    C_{1}^{*}e^{-i\delta}\\
  C_{2}^{*} & t_{a} & B_{3}^{*} & A_{4} & \cdots & 0 & 0 & 0 \\
  \vdots& \vdots& \vdots& \vdots& \ddots& \vdots& \vdots& \vdots  \\
  C_{M-1}e^{i\delta} & 0 & 0 & 0 & \cdots&A_{M-2} & B_{M-2} & t_{a} \\
  t_{a}e^{i\delta} & C_{M}e^{i\delta} & 0 & 0 & \cdots & B_{M-2}^{*} &
    A_{M-1} & B_{M-1} \\
  B_{M}e^{i\delta} & t_{a}e^{i\delta} & C_{1}e^{i\delta} & 0 & \cdots
    & t_{a} & B_{M-1}^{*} & A_{M}
  \end{array}\right)                                \label{eq:9}
\end{equation}
with $\delta=k_{x}qa/2$ ($k_{x}qa$) for an even (odd) $p$.
\begin{multicols}{2}

\begin{figure}
\centerline{\epsfig{figure=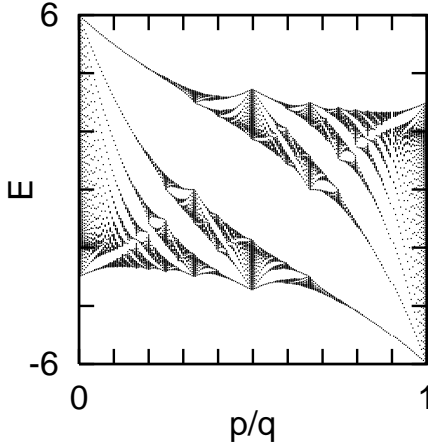,width=0.65\linewidth}}
\narrowtext
\caption{Energy eigenvalues versus $\phi$ for a triangular
lattice with isotropic NN hopping integrals ($t_{a}=t_{b}=t_{c}=1$).
Calculations are performed for $\phi=p/199$ with $1\le p\le 198$, and
the central value of $\vec{k}$ in the MBZ is taken into account.}
\end{figure}

\section{Numerical Results and Discussion}

Before presenting our numerical results, we would like to point out the
mistake made by Gumbs and Fekete in Ref.~24. In converting
Eq.~(\ref{eq:10}) into Eq.~(\ref{eq:0}) [i.e., Eq.~(2) of Ref.~24],
they made a mistake in operating $e^{\pm
i(p_{x}\pm\sqrt{3}eBx)a/2\hbar}$ to $\Psi_{m,n}$. According to them,
the results of the operations are given by
\begin{eqnarray}
 e^{\pm i(p_{x}+\sqrt{3}eBx)a/2\hbar}\Psi_{m,n}
 =e^{\pm i\pi\phi m}\Psi_{m\mp 1,n},    \nonumber \\
 e^{\pm i(p_{x}-\sqrt{3}eBx)a/2\hbar}\Psi_{m,n}
 =e^{\mp i\pi\phi m}\Psi_{m\mp 1,n}.            \label{eq:11}
\end{eqnarray}
However, these are incorrect since the fundamental commutation relation
$[x,p_{x}]=i\hbar$ is ignored in the operation. The correct results
taking into account this relation should be expressed as\cite{22}
\begin{eqnarray}
 e^{\pm i(p_{x}+\sqrt{3}eBx)a/2\hbar}\Psi_{m,n}
 =e^{\pm i\pi\phi(m\mp 1/2)}\Psi_{m\mp 1,n}, \nonumber \\
  e^{\pm i(p_{x}-\sqrt{3}eBx)a/2\hbar}\Psi_{m,n}
 =e^{\mp i\pi\phi(m\mp 1/2)}\Psi_{m\mp 1,n}.      \label{eq:12}
\end{eqnarray}
We can easily check that the same equation as Eq.~(\ref{eq:4}) with
$t_{a}'=t_{b}'=t_{c}'=0$ can be obtained by the method of the Peierls
substitution if Eq.~(\ref{eq:12}) instead of Eq.~(\ref{eq:11}) is
applied. In fact, Eq.~(6) of Ref.~24, which was derived by means of
Eq.~(\ref{eq:11}), is a non-Hermitian matrix, and thus it is
unreasonable to obtain real energy eigenvalues from this matrix.

Figure~1 shows the $E-\phi$ diagram obtained by diagonalizing
Eq.~(\ref{eq:9}) for the parameters $t_{a}=t_{b}=t_{c}=1$, $\vec{k}=0$,
and $\phi=p/199$ with $1\le p\le 198$ in the absence of the NNN hopping
integral ($t_{a}'=t_{b}'= t_{c}'=0$). It is obvious that the energy
spectrum is {\it not} symmetric about $\phi=1/2$ and there is {\it no}
distinct array of energy eigenvalues parallel to the energy axis in the
two largest energy gaps, contrary to the argument of Gumbs and Fekete
[see

\begin{figure}
\narrowtext
\centerline{\epsfig{figure=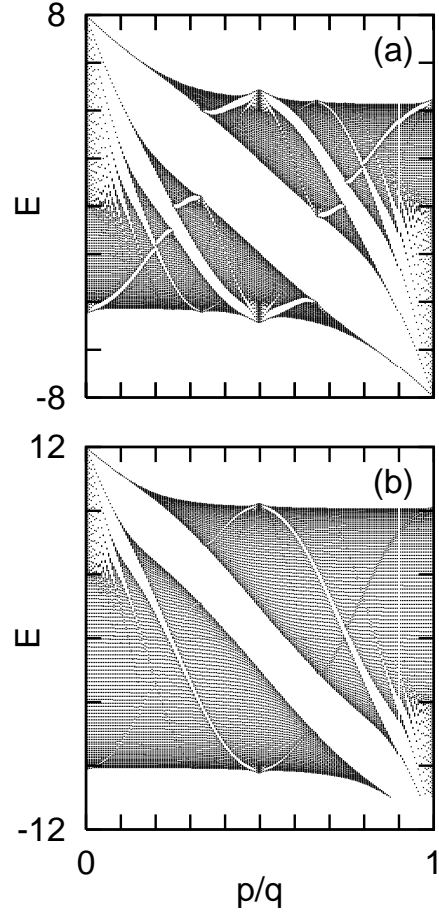,width=0.65\linewidth}}
\caption{Energy eigenvalues versus $\phi$ for a lattice with (a)
$(t_{a},t_{b},t_{c})=(2,1,1)$ and (b) $(t_{a},t_{b},t_{c})=(4,1,1)$.
Other parameters are the same as Fig.~1.}
\end{figure}

\noindent
Fig.~1 and the relevant arguments of Ref.~24]. In fact, Fig.~1 of
this paper is exactly the same as Fig.~3 of Ref.~21 presented by Claro
and Wannier if $E$ is replaced by $-(E-\epsilon_{0})/2\epsilon_{1}$,
which is quite natural because both of the two papers deal with the
same problem. The only difference between the two papers lies in
choosing the $x$ and $y$ directions; the axes taken in Ref.~21 are
turned by $\pi/6$ from those in this paper. Note that, even though the
choice of the axes in Ref.~21 yields a three-term difference equation
[see Eq.~(10) of Ref.~21] while the choice in this paper yields a
five-term difference equation in the absence of NNN hopping integrals,
the resultant energy band structures are the same, as shown above.

\subsection{Effects of anisotropic NN hopping integrals}

Figure~2 is the $E-\phi$ diagram for the parameters $t_{b}=t_{c}=1$,
$\vec{k}=0$, and (a) $t_{a}=2$ and (b) $t_{a}=4$ in the absence of NNN
hopping integrals. The increase of total bandwidths and the occurrence
of gap closing except for a few large ones are clearly seen. We argue
that these behaviors may be generic effects of the hopping anisotropy,
since the same phenomena were also observed in the square
lattice.\cite{13}$^{-}$\cite{15} Note that Fig.~3 of Ref.~24 is the
$E-\phi$ diagram plotted for the same parameters used in plotting
Fig.~2(a) of this paper, from which Gumbs and Fekete argued that the
bottom of the energy band is very flat near $E=-4$ and that the
spectrum is symmetric about $\phi=1/2$. However, it is evident that
their arguments are incorrect. In regard to Figs.~2 $-$ 4 of Ref.~24,
we would like to point out that the $E-\phi$ diagrams for the
parameters $(t_{a},t_{b},t_{c})=(1,2,1)$, $(2,1,1)$, and $(1,1,2)$
should be exactly the same when all the values of $\vec{k}$ in the
magnetic Brillouin zone (MBZ; $|k_{x}|\le 2\pi/Ma$ and
$|k_{y}|\le\pi/\sqrt{3}a$) are taken into account, which means that
there is no preferable direction in the lattice plane under a uniform
perpendicular magnetic field.

\begin{figure}
\narrowtext
\centerline{\epsfig{figure=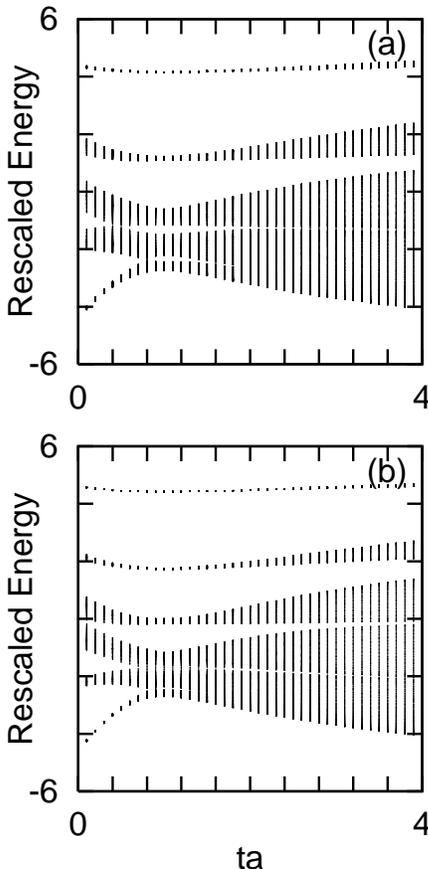,width=0.65\linewidth}}
\caption{Rescaled energy eigenvalues as a function of
$t_{a}$ for a lattice with (a) $\phi=1/5$ and (b) $\phi=1/6$, and
$t_{b}=t_{c}=1$. All the values of $\vec{k}$ in the MBZ are taken into
account.}
\end{figure}

In order to get more information on the anisotropy dependence of the
subgap widths, we calculate the band structure as a function of the
hopping anisotropy. Note that the energy eigenvalues in the absence of
a magnetic field satisfy the condition $|\varepsilon(\vec{k})|\le
2(t_{a}+t_{b}+t_{c})$ [see Eq.~(\ref{eq:10})] and the total bandwidth
increases linearly with the increase of $t_{a}$. Assuming the situation
also holds in the presence of a uniform magnetic field, we can
introduce the rescaled energy defined by
\begin{equation}
 E_{re}=3E/(t_{a}+t_{b}+t_{c}).                 \label{eq:13}
\end{equation}
Then, $E_{re}$ is always in the range of $[-6,6]$ regardless of the
strength of the hopping anisotropy, and we can examine more clearly the
dependence of the subgap widths on the hopping anisotropy, in
comparison with the isotropic case.

Figure~3(a) shows the $t_{a}$ dependence of the band structure for
$\phi=1/5$, where $t_{b}=t_{c}=1$ and all the values of $\vec{k}$ in
the MBZ are taken into account. At isotropic case ($t_{a}=1$), there
are five subbands and four subgaps. However, introducing the hopping
anisotropy ($t_{a}\neq 1$) changes the band structure considerably. Let
us number the subgaps from the bottom of the energy spectrum. Then, in
the regime of $t_{a}>1$, the first and the second subgaps close for a
small amount of anisotropy, while the third and the fourth subgaps
survive even for large values of anisotropy. However, since the widths
of the third and the fourth subgaps decrease with increasing $t_{a}$,
we expect that these subgaps will also close in the limit of
$t_{a}\rightarrow\infty$. And, in the regime of $t_{a}<1$, the widths
of the third and the fourth subgaps decrease with decreasing $t_{a}$,
as in the case of $t_{a}>1$. Meanwhile, the first and the second
subgaps exhibit an interesting behavior. As $t_{a}$ decreases, the
width of the first (second) subgap decreases such that it closes at
$t_{a}\simeq 0.8$ ($t_{a}\simeq 0.45$). And, as $t_{a}$ decreases
further, the subgap reopens and its width increases such that the width
of the first (second) subgap becomes the same as that of the fourth
(third) subgap, resulting in a symmetric band structure about $E=0$ in
the limit of $t_{a}\rightarrow 0$. Since the triangular lattice with
the parameters $(t_{a},t_{b},t_{c})= (0,1,1)$ is topologically
equivalent to the square lattice with isotropic NN hopping integrals,
it is natural of the band structure to be symmetric about $E=0$.

As an another example, we plot in Fig.~3(b) the $t_{a}$ dependence of
the band structure for $\phi=1/6$. At isotropic case ($t_{a}=1$), there
might be six subbands and five subgaps. However, due to the occurrence
of band touching between the second and the third subbands, the second
subgap is missing and the energy spectrum consists of four subgaps.
And, in the regime of $t_{a}>1$, we observe that the first subgap close
even for a small amount of anisotropy and the third subgap close for a
large value of anisotropy. Besides, since the widths of the fourth and
the fifth subgaps decrease with increasing $t_{a}$, the closing of
these subgaps is expected in the limit of $t_{a}\rightarrow\infty$.
Meanwhile, the $t_{a}$ dependence of the missing second subgap exhibits
an interesting feature; the subgap opens with increasing $t_{a}$, even
though its width is very small. In the regime of $t_{a}<1$, we can see
that the width of the first (second) subgap increases with decreasing
$t_{a}$ such that the width becomes the same as that of the fifth
(fourth) subgap, resulting in a symmetric band structure about $E=0$ in
the limit of $t_{a}\rightarrow 0$. And, the width of the third subgap
decreases with decreasing $t_{a}$ and closes at $t_{a}=0$. Thus, there
occurs a degeneracy at the band center of the energy spectrum, as in
the case of the square lattice with isotropic NN hopping integrals.

\end{multicols}

\begin{figure}
\widetext
\centerline{\epsfig{figure=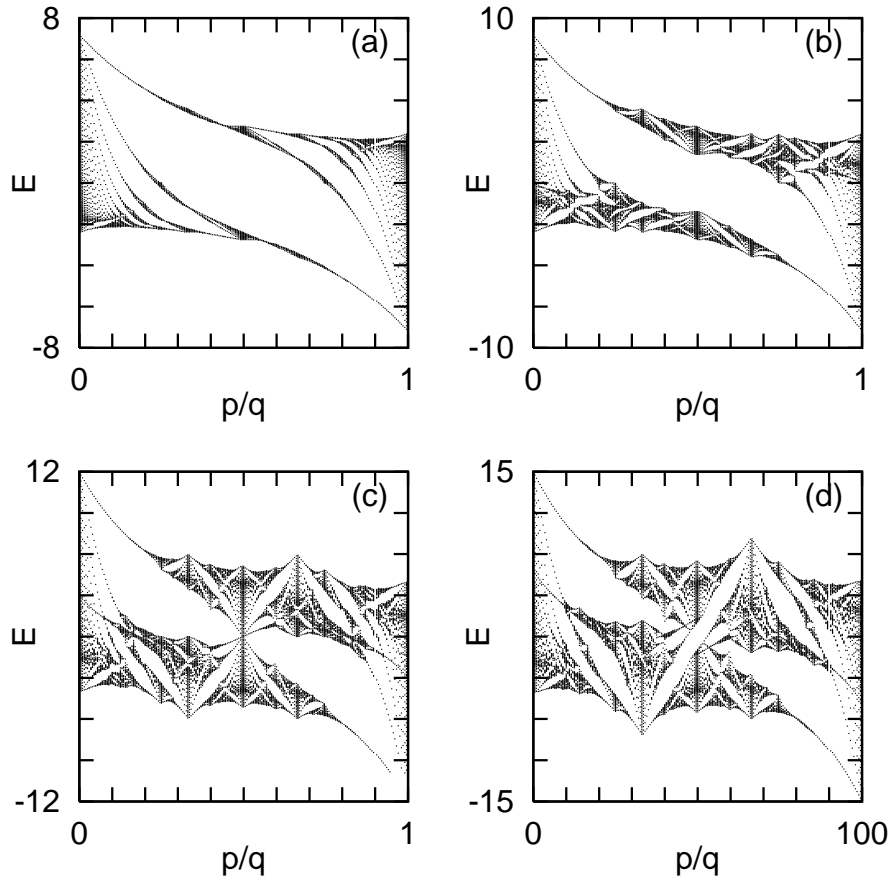,width=0.65\linewidth}}
\caption{Energy eigenvalues versus $\phi$ for a lattice with
(a) $t'=0.2$, (b) $t'=0.5$, (c) $t'=1.0$, and (d) $t'=1.2$,
respectively. Other parameters are the same as Fig.~1.}
\end{figure}

\begin{multicols}{2}

\subsection{Effects of isotropic NNN hopping integrals}

In calculating energy eigenvalues with varying the NNN hopping
integrals, we set $t_{a}=t_{b}=t_{c}=1$ and assume that the NNN hopping
integrals are isotropic, i.e., $t_{a}'=t_{b}'=t_{c}'(\equiv t')$.
Figure~4 shows the $E-\phi$ diagram for (a) $t'=0.2$, (b) $t'=0.5$, (c)
$t'=1.0$, and (d) $t'=1.2$, respectively. We can see that the fractal
structure in the energy spectrum appears despite of introducing the NNN
hopping integral, which may attribute to the isotropy of the NN and the
NNN hopping integrals. We can also see that, even though the total
bandwidths for $\phi=0$ and $1$ increase monotonically with increasing
$t'$, the bandwidths for generic values of $\phi$ depend crucially on
$t'$, resulting in diverse shapes of $E-\phi$ diagrams. For small
values of $t'$, the bandwidths become narrower than the case of $t'=0$
and the resultant diagram exhibits a vary narrow structure [see
Fig.~4(a)]. However, the bandwidths increase with increasing $t'$,
resulting in a fat structure of $E-\phi$ diagram [see Figs.~4(b) -
4(d)].

Figure~5 shows the $t'$ dependence of the band structure for
$\phi=1/2$, which exhibits two critical values of $t'$. The subgap
width increases with increasing $t'$ up to $t_{1}'\simeq 0.25$, and
decreases up to $t_{2}'=1$ such that gap closing occurs at $t_{2}'$.
And, then it increases again for $t'>t_{2}'$. It may be interesting to
compare this behavior with the case of the square lattice, since the
triangular lattice with $t'\neq 0$ is topologically equivalent to the
square lattice with isotropic (anisotropic) NNN hopping integrals if
$t'=1$ ($t'\neq 1$). In the square lattice, introducing the NNN hopping
integrals removes the degeneracy appearing at $E=0$ for even values of
$q$.\cite{16} However, in the triangular lattice, a degeneracy occurs
at $E=0$ in the energy spectrum for $\phi=1/2$ when $t'=t$.

The $t'$ dependence of the band structures for $\phi=1/3$ and $2/3$
also exhibit interesting features. In these cases, there are three
subgaps at $t'=0$. And, with increasing $t'$, the widths of these
subbands decrease up to a certain value of $t'$ [see Fig.~4(a)], and
then increase such that the two subgaps close [see Figs.~4(b)-4(d)] for
large values of $t'$. Particularly, the lowest (highest) subband edge
for $\phi=1/3$ (2/3) goes downward (upward) with increasing

\begin{figure}
\narrowtext
\centerline{\epsfig{figure=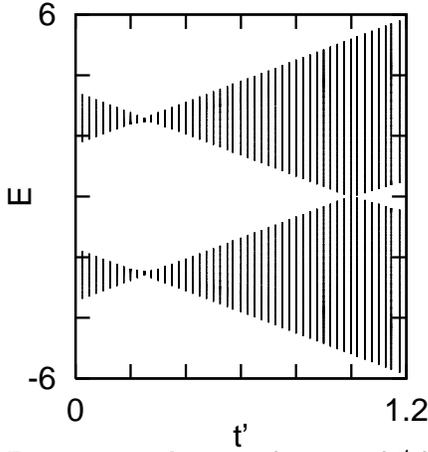,width=0.65\linewidth}}
\caption{Energy eigenvalues as a function of $t'$ for a
lattice with $\phi=1/2$. All the values of $\vec{k}$ in the MBZ are
taken into account.}
\end{figure}

\begin{figure}
\narrowtext
\centerline{\epsfig{figure=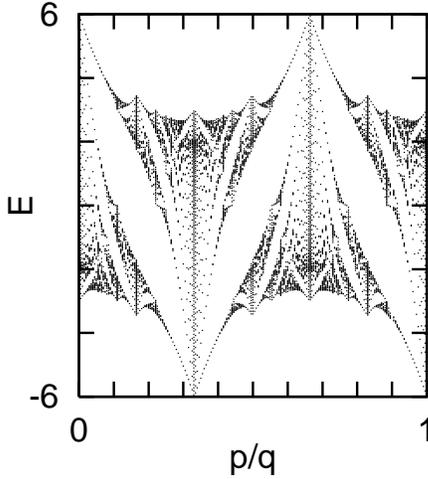,width=0.65\linewidth}}
\caption{Energy eigenvalues versus $\phi$ for a lattice with
$t=0$ and $t'=1$. The data are obtained for $\phi=p/199$ with $1\le
p\le 198$ and for $\vec{k}=0$.}
\end{figure}

\noindent $t'$ such
that $E_{l}(\phi=1/3)=E_{l}(\phi=1)$ [$E_{h}(\phi=2/3)=E_{h}(\phi=0)$]
in the limit of $t'/t\rightarrow \infty$, where $E_{l (h)}$ means the
lowest (highest) subband edge. From this behavior, it can be expected
that the periodicity of the $E-\phi$ diagram in the limit of
$t'/t\rightarrow\infty$ reduces by three times in $\phi$. Figure~6
shows the $E-\phi$ diagrams for the lattice with $t=0$ and $t'=1$,
which confirms this expectation. Note that it is nothing but an energy
spectrum for the lattice with the isotropic hopping integral ($t'=1$)
and the lattice constant $\sqrt{3}a$. In order to see the $t'$
dependence of the band structure for generic values of $\phi$, we plot
in Fig.~7 the $E-t'$ diagrams for (a) $\phi=1/7$ and (b) $\phi=2/7$. We
can see that gap closing, gap reopening, and band crossing occur
depending on the values of $t'$.

Before concluding this section, we discuss the symmetry of the energy
spectrum. Denoting the set of energy eigenvalues for a wave vector
$\vec{k}$ under a flux $\phi$ as $E(\phi; \vec{k})$ and the energy
spectrum taking into account all the val-

\begin{figure}
\narrowtext
\centerline{\epsfig{figure=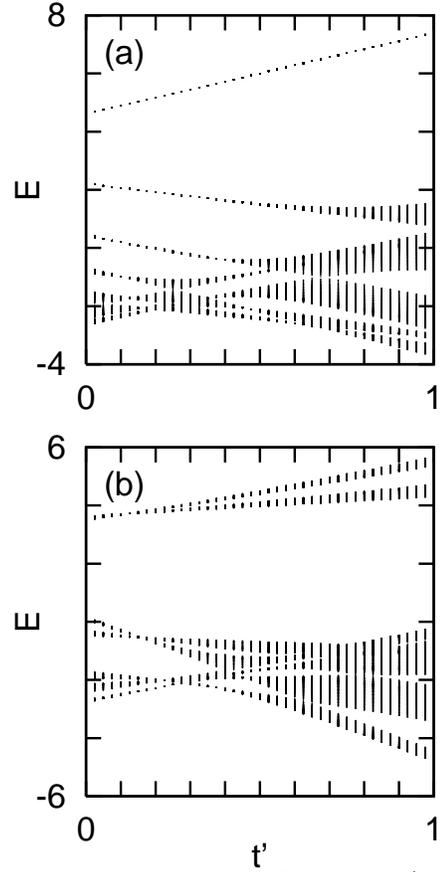,width=0.65\linewidth}}
\caption{Energy eigenvalues as a function of $t'$ for a
lattice with (a) $\phi=1/7$ and (b) $\phi=2/7$. All the values of
$\vec{k}$ in the MBZ are taken into account.}
\end{figure}

\noindent ues of $\vec{k}$ in the MBZ as
$E(\phi)$, we can see the following properties of the band structure
for generic values of NN hopping integrals: (i)
$E(\phi;\vec{k})=E(\phi;-\vec{k})$, (ii) $E(\phi)=E(-\phi)$, (iii)
$E(\phi)=-E(1-\phi)$ with $0\le \phi\le 1/2$, and (iv)
$E(\phi)=E(\phi+2)$. Note that the property (i) can be easily checked
in Eq.~(\ref{eq:9}); ${\bf A}\rightarrow {\bf A}^{\dagger}$ under the
transformation $\vec{k}\rightarrow -\vec{k}$. Note also that the
property (ii) means that there is no way for an electron to discern the
direction of the magnetic field. Figures~1 and 2 show that the
properties (iii) and (iv) hold in the presence of the NN hopping
anisotropy. And, Fig.~4 also shows that the properties (i)--(iv) hold
in the presence of NNN hopping integrals except for the case of $t=0$;
when $t=0$, the properties (iii) and (iv) are replaced by
$E(\phi)=-E(1/3-\phi)$ with $0\le \phi\le 1/6$ and
$E(\phi)=E(\phi+2/3)$, respectively, as can be seen in Fig.~6.

\section{Summary}

In summary, the effects of anisotropic NN hopping integrals and of
isotropic NNN hopping integrals on the magnetic subband structure of an
electron on a triangular lattice were studied within the tight-binding
approximation. It was found that the phenomena of bandwidth broadening,
gap closing, and gap opening are generic feature of introducing the NN
hopping anisotropy. It was also found that the magnetic subband
structure changes considerably by introducing the NNN hopping
integrals; bandwidth broadening, gap closing, gap reopening, and band
crossing occur depending on the strength of the isotropic NNN hopping
integrals. In the study, the dependence of the subgap widths on the
hopping anisotropy and the NNN hopping integrals was illustrated in
detail, and the symmetries of the magnetic subband structures with and
without the hopping anisotropy and the NNN hopping integrals were also
discussed. In order to deepen the understanding on the band structure
of the triangular lattice, a further study of the effect of, for
example, nonuniform magnetic fields\cite{23} and an interaction between
electrons\cite{24} on the magnetic subband structure of the lattice is
required, and the work is on doing.

\acknowledgments{The author is grateful to Dr. J. Chung and Mrs. B. K.
Min for stimulating discussions.}

\end{multicols}

\begin{references}
\bibitem[*]{0} Electronic address: ogy@hnu.hankyong.ac.kr
\bibitem{1}
  R. Peierls, Z. Phys. {\bf 80}, 763 (1933);
  P. G. Harper, Proc. Phys. Soc. London A {\bf 68}, 874 (1955);
  W. Kohn, Phys. Rev. {\bf 115}, 1460 (1959);
  M. Ya Azbel, Zh. \'{E}ksp. Teor. Fiz. {\bf 46}, 929 (1994) [Sov.
     Phys. JETP {\bf 19}, 634 (1964)].
\bibitem{2}
  D. R. Hofstadter, Phys.  Rev. B {\bf 14}, 2239 (1976).
\bibitem{3}
  G. H. Wannier, Phys. Status Solidi B {\bf 88}, 757 (1978);
  G. H. Wannier, G. M. Obermaier, and R. Ray, {\it ibid.} {\bf 93},
     337 (1979).
\bibitem{4}
  A. H. MacDonald, Phys. Rev. B {\bf 28}, 6713 (1983).
\bibitem{5}
  J. B. Sokoloff, Phys. Rep. {\bf 126}, 189 (1985).
\bibitem{6}
  M-C. Chang and Q. Niu, Phys. Rev. B {\bf 53}, 7010 (1996).
\bibitem{7}
  Y. Hatsugai, M. Kohmoto, and Y-S. Wu, Phys. Rev. B {\bf 53}, 9697
    (1996).
\bibitem{a}
 D. J. Thouless, M. Kohmoto, P. Nightingale, and M. Den. Nijs, Phys.
    Rev. Lett. {\bf 49}, 405 (1982);
 R. Rammal, G. Toulouse, M. T. Jaekel, and B. I. Halperin, Phys.
    Rev. B {\bf 27}, 5142 (1983).
\bibitem{b}
 P. W. Anderson, Phys. Scr. T {\bf 27}, 60 (1989);
 P. Lederer, D. Poilblanc, and T. M. Rice, Phys. Rev. Lett. {\bf 63},
    1519 (1989).
\bibitem{c}
 S. Alexander, Phys. Rev. B {\bf 27}, 1541 (1983);
 R. Rammal, T. C. Lubensky and G. Toulose, {\it ibid}. {\bf 27},
    2820 (1983);
 Q. Niu and F. Nori, {\it ibid}. {\bf 39}, 2134 (1989).
\bibitem{8}
  B. Pannetier, J. Chaussy, R. Rammal, and J-C. Villegier, Phys. Rev.
     Lett. {\bf 53}, 1845 (1984).
\bibitem{9}
  D. Weiss, M. L. Roukes, A. Menshig, P. Grambow, K. von Klitzing,
     and G. Wieman, Phys. Rev. Lett. {\bf 66}, 2790 (1991).
\bibitem{10}
  A. Lorke, J. P. Kotthaus, and K. Ploog, Phys. Rev. B {\bf 44},
     3447 (1991).
\bibitem{11}
  R. R. Gerhardts, D. Weiss, and U. Wulf, Phys. Rev. B {\bf 43}, 5192
     (1991);
  R. R. Gerhardts and D. Pfannkuche, Surf. Sci. {\bf 263}, 324 (1992).
\bibitem{12}
  M. Kohmoto, Phys. Rev. B {\bf 39}, 11943 (1989).
\bibitem{13}
  Y. Hasegawa, Y. Hatsugai, M. Kohmoto, and G. Montambaux, Phys. Rev. B
     {\bf 41}, 9174 (1990).
\bibitem{14}
  S. N. Sun and J. P. Ralston, Phys. Rev. B {\bf 44}, 13603 (1991).
\bibitem{15}
  G. Y. Oh, J. Jang, and M. H. Lee, J. Korean Phys. Soc. {\bf 29}, 261
    (1996).
\bibitem{16}
  Y. Hatsugai and M. Kohmoto, Phys. Rev. B {\bf 42}, 8282 (1990);
  J. H. Han, D. J. Thouless, H. Hiramoto, and M. Kohmoto, Phys. Rev. B
  {\bf 50}, 11365 (1994).
\bibitem{17}
  D. Langbein, Phys. Rev. {\bf 180}, 633 (1969).
\bibitem{18}
  F. H. Claro and G. H. Wannier, Phys. Rev. B {\bf 19}, 6068 (1979).
\bibitem{19}
  D. J. Thouless, Phys. Rev. B {\bf 28}, 4272 (1983).
\bibitem{20}
  O. K\"{u}hn, P. E. Shlbmann, V. Fessatidis, and H. L. Cui, J. Phys.:
    Condens. Matter {\bf 5}, 8225 (1993).
\bibitem{21}
  G. Gumbs and P. Fekete, Phys. Rev. B {\bf 56}, 3787 (1997).
\bibitem{22}
  S. Gasiorowicz, {\it Quantum Physics} (John Wiley \& Sons Inc.,
  1974), p. 137.
\bibitem{23}
  A. Barelli, J. Bellissard, and R. Rammal, J. Phys. (Paris) {\bf 51},
  2167 (1990);
  G. Y. Oh and M. H. Lee, Phys. Rev. B {\bf 53}, 1225 (1996);
  Q. W. Shi and K. Y. Szeto, {\it ibid.} {\bf 56}, 9251 (1997);
  G. Y. Oh, {\it ibid.} {\bf 60} (in press).
\bibitem{24}
 H. Doh and S-H. Suck Salk, Phys. Rev. B {\bf 57}, 1312 (1998).

\end{references}
\end{document}